\documentclass[%
 reprint,
 unsortedaddress,
groupedaddress,
 amsmath,amssymb,
 aps,
 prper,
]{revtex4-2}

\usepackage{url}
\usepackage{hyperref}
\hypersetup{breaklinks=true}

\usepackage{graphicx}
\usepackage{dcolumn}
\usepackage{bm}
\usepackage{booktabs}
\usepackage{verbatim}

\usepackage{enumerate}
\usepackage[shortlabels]{enumitem}
\usepackage{multirow}

\usepackage{xspace}
\newcommand{\eclass}{E-CLASS\xspace}

\newcommand{\see}{Self-Evaluated Expertise\xspace}
\newcommand{\See}{Self-Evaluated Expertise\xspace}

\begin{document}

\preprint{APS/123-QED}

\title{Self-Evaluated Expertise in experimental physics:\\ a measure of students' physics self-recognition}

\author{Michael F. J. Fox}
\email{michael.fox@imperial.ac.uk}
\affiliation{%
 Department of Physics, Imperial College London, Prince Consort Road, London, SW7 2AZ, UK
}%

\author{Taylor O. Pomfret}
\affiliation{%
 Department of Physics, Imperial College London, Prince Consort Road, London, SW7 2AZ, UK
}%

\date{\today}

\begin{abstract}
We introduce and theoretically justify a new measure of the self-recognition component of student physics identity called Self-Evaluated Expertise (SEE). This measure is constructed such that it can be extracted from existing responses to the E-CLASS. In this work, we compare scores from SEE with the traditional measure calculated from the E-CLASS, which probes student views about experimental physics, to show that the SEE score is a quantitatively different measure. Consequently, we show that student self-recognition decreases from pre-instruction administration of the E-CLASS to the post-instruction administration when averaged across data from 494 courses having taken place between 2016--2019.
\end{abstract}

\maketitle

\section{Introduction}\label{sec:intro}
One common goal of physics education is to develop students' views to become more expert like~\cite{etkina2010,Chi1981,Zwickl2014}. It has been argued that an expert in a subject demonstrates a distinct epistemology from that of novices~\cite{Sin2014, Taasoobshirazi08, petcovic2007} and as such their answers to survey questions can be used as the yardstick for measuring how close students are to thinking like experts~\cite{Zwickl2014,Adams2006class}. Notwithstanding the issue of establishing consensus among experts~\cite{Zwickl2014}, this method has been used to score epistemological surveys such as the Colorado Learning Attitudes about Science Survey (CLASS)~\cite{Adams2006class} and the Colorado Learning Attitudes about Science Survey for Experimental Physics (E-CLASS)~\cite{Zwickl2014}. This definition of what is an expert is external to the student and therefore \textit{may} be interpreted as providing an objective measure of epistemological development. 

Another possible measure of epistemological development is how students view themselves relative to their own perception of what an expert thinks. This centres the student in the analysis. This is naturally a more complex measure, as students' view of their own thinking and their view of the way experts think could change over time. Let us consider the case where the goal of a course is for students to think of themselves as experts, this means that they would say that an expert would give the same response to a survey question as they do. The obvious issue is that this rewards students who are confident in their own abilities even if those are misplaced relative to what a ``true'' expert would think - hence it measures a student's calibration of their own abilities. With this example we have illustrated that such a measure is recording something rather different from traditional scoring for epistemological surveys. In this paper we introduce a measure we call Self-Evaluated Expertise (SEE), and from the outset we relate it to an aspect of students' identity.
 
Student identity development frameworks have been used extensively in the discourse of the experiences of under-represented groups in physics~\cite{carlone2007understanding,hazari_connecting_2010,hyater-adams_critical_2018}. We base this work in the Critical Physics Identity framework of Hyater-Adams et al.~\cite{hyater-adams_critical_2018}, which constructs identity through six components: three related to racialized identity resources (relational, ideational, and material), and three related to physics identity constructs (recognition, interest, and competence/performance). We particularly focus on the recognition component, which Hyater-Adams et al. break down into internal and external sources of recognition (and positive and negative for encouraging participation in physics). The distinction between internal and external recognition builds on Carlone and Johnson's science identity framework~\cite{carlone2007understanding}, who identified self-recognition and recognition by others as two components of recognition. It is this self-recognition component of identity that we relate our Self-Evaluated Expertise measure to. The assumption of this paper is that if one believes that they are an expert --- or at least hold the same views as an expert --- this contributes positively to confidence and participation in the discipline of physics and has important consequences for who considers themself capable of becoming a physicist or being a ``physics person''~\cite{hyater-adams_critical_2018}.  

Typically works that investigate aspects of identity use interview and/or survey methodologies to collect data, however, interviews are limited, typically, to small sample sizes, and the practicalities of running multiple surveys needed to probe different aspects of students' development on a course can lead to survey fatigue. This motivates our interest in developing an analysis of an existing survey (\eclass) in a new direction that we hope to show gives insights into students' physics identities.

Indeed, previous work with \eclass has shown that students often have a good idea of what an expert response to the survey items would be, even if they do not hold those views themself and that the gap between these two decreases when comparing first-year and beyond-first-year students~\cite{Wilcox2017}. Similar work by Gray et al.~\cite{Gray08} using the CLASS survey compared ``physicist'' (expert) scores and ``personal'' (student's own scores) and investigated how the difference between the two varied with gender. They found:
\begin{quote}
     Women have a larger gap between their ``physicist'' and ``personal'' scores... on 78\% of the scored statements. Six of the seven statements that had gap differences greater than +10\% had men’s ``personal'' scores that were notably more expertlike than the women’s ``personal'' scores... These six statements... all dealt with either the student’s interest in physics or the student’s confidence in his or her problem solving abilities.
\end{quote}
This, therefore, motivates the development of a measure that captures such differences to help explore the identity-related factors that affect engagement with (experimental) physics.

The purposes of this paper are two-fold: 
\begin{enumerate}[start=1,label={\bfseries P\arabic*:}]
    \item theoretically justify and practically describe the construction of a measure of self-recognition using the \eclass survey, and
    \item demonstrate that measure provides new information about students that would not be accessible using the traditional \eclass scores.
\end{enumerate}

We approach the first purpose by grounding the work in the physics identity literature above in this Introduction, then, in the following sections, defining the Self-Evaluated Expertise score and explaining how it can be considered a measure of self-recognition. To meet the second purpose we apply the measure to data collected with \eclass and compare the score with the traditional scoring system used to analyse \eclass (the YOU score, see Sec.~\ref{sec:scoring-eclass}), highlighting results that provide new insights into this one component of student identity. Practically, we demonstrate the differences between the SEE measure and the traditional \eclass scoring by: (1) checking the correlation between \eclass YOU and SEE scores total scores; (2) checking the correlation between YOU and SEE mean item scores; and, (3) investigating whether \see scores change in the same way as \eclass scores from the beginning to the end of a course on an item-by-item basis. Throughout this process, we aim to provide the reader insight into how to interpret SEE scores. While part of the motivation comes from the gender-based observations made by Gray et al.~\cite{Gray08}, we leave the exploration of how the SEE score provides insight into demographic inequities to later work.

\section{Methods}\label{sec:methods}
\subsection{Construction of the SEE score}
\subsubsection{The \eclass}\label{sec:methods:eclass-intro}
\begin{figure*}
    \centering
    \includegraphics[width=0.8\linewidth]{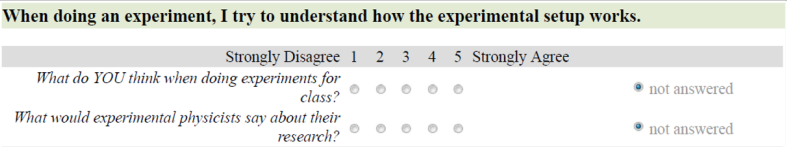}
    \caption{Item 1 from the \eclass pre-test. Reproduced from the \eclass website with permission~\cite{eclass-website}.}
    \label{fig:example-item}
\end{figure*}
The \eclass principally consists of 30 items, each containing a statement associated with typical learning goals of undergraduate physics labs, with some items particularly targeting student affect and confidence when doing an experiment~\cite{Zwickl2014, Wilcox16}. Within each item there is a statement and two prompts: ``What do YOU think when doing experiments in class?" (YOU question) and ``What would experimental physicists say about their research?" (Expert question). For each prompt respondents are given a 5-point scale labelled with integers 1 to 5 and with 1 noted to mean ``Strongly Disagree'' and 5 noted to mean ``Strongly Agree''. The layout of these questions within an item is illustrated in Figure~\ref{fig:example-item}.

Respondents are asked to fill in the survey at the start and end of their lab course: which will be referred to as pre-test and post-test respectively. Respondents to the post-test are given an additional set of questions, which asks for details regarding their course (e.g., declared major, length of their study, level of interest in physics, future plans) and also their demographics (e.g., gender and race/ethnicity). While the focus of this work is not on demographic effects, we follow the recommendations of Kanim and Cid~\cite{kanim2020} and report demographic data in Appendix~\ref{sec:demographics}.

\subsubsection{Scoring of the \eclass}\label{sec:scoring-eclass}
\begin{figure}[ht] 
    \centering
    \includegraphics[width=0.8\linewidth]{"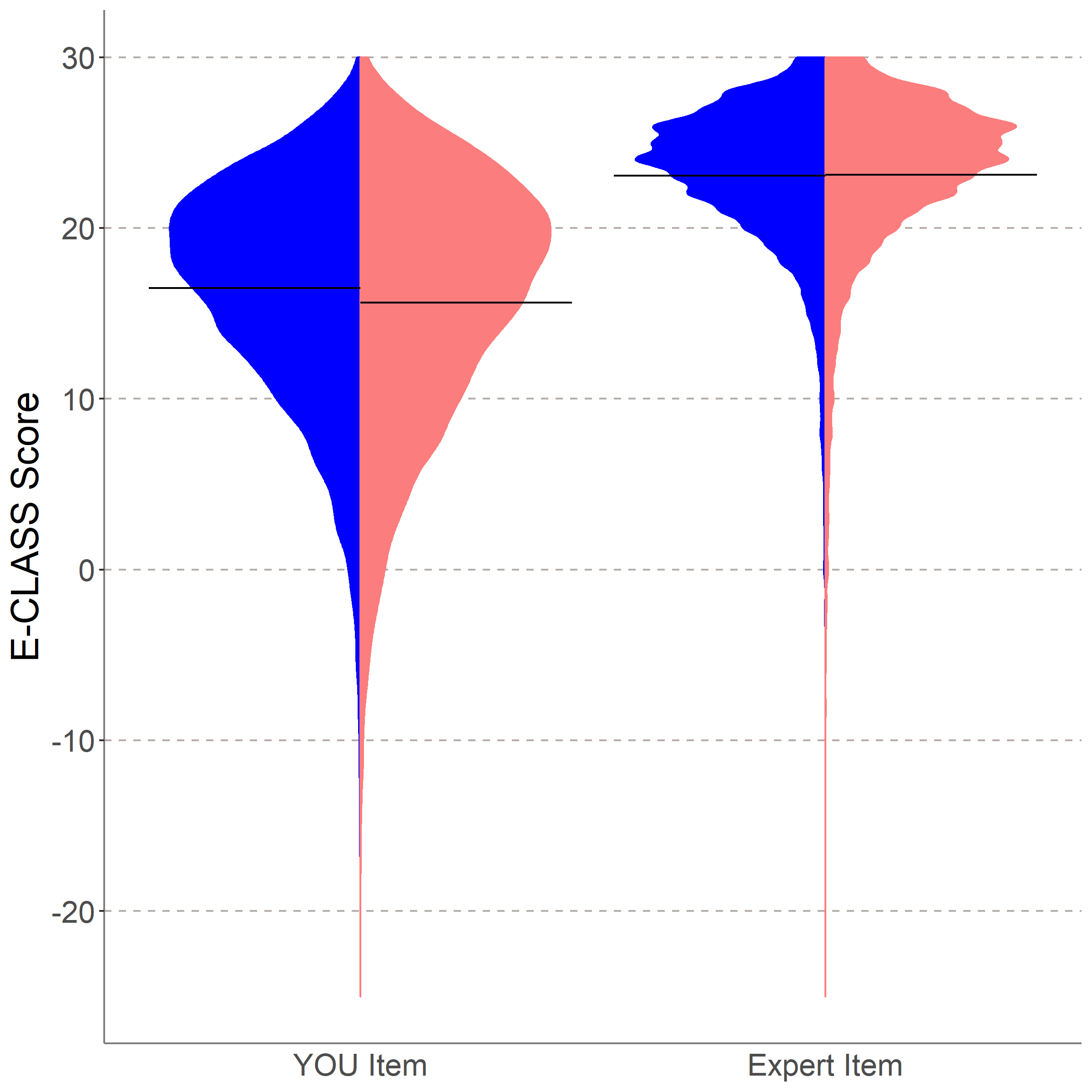"}
    \caption{Two-sided violin plots showing the breakdown of the overall E-CLASS scores of the respective YOU and Expert items is shown in the figure, with the blue side of the violin plot depicting the pre-test response, and the red side depicting the post-test response. Horizontal black lines indicate the mean YOU and Expert score as described in Section~\ref{sec:scoring-eclass}. Expert score pre-test 23.06 $\pm$ 0.03 and post-test 23.11 $\pm$ 0.04. YOU score pre-test 16.49 $\pm$ 0.05 and post-test 15.63 $\pm$ 0.05. Uncertainties are standard errors on the mean.}
    \label{fig:eclass_violins}
\end{figure}

An individual's \eclass score is obtained by collapsing the 5-point scale to a 3-point scale where responses are coded as either agree (values 4 and 5), disagree (values 1 and 2), or remaining neutral (value 3) with the item statement. The reason for this is to mitigate ambiguity between the respondent's interpretation of the extremes. These responses are then compared to an expert-like consensus gathered from professional experimental physicists (see Table I in Ref.~\cite{Zwickl2014}). If the response matches that of the consensus of experts (e.g., both agree with the item statement), a score of +1 is given for the item; if the response directly opposes that of the expert (e.g., the respondent disagrees with the item statement while the expert consensus is to agrees with it), a score of -1 is given for the item; and if the response is neutral, a score of 0 is given for the item. The total E-CLASS score is then given by the sum of the scores from the 30 individual items. Scores exist in the interval [-30,30], see Figure~\ref{fig:eclass_violins}. As we will be introducing a new scoring system for the \eclass responses, it is pertinent to introduce some terminology. Therefore, from now on we will refer to this traditional scoring system as the YOU score, as it is derived from the responses to the YOU question only. Most studies using the \eclass have analysed results generated from the YOU scores~\cite{wilcox2018summary,Fox21}.

One could argue that the YOU score is an objective measure as it is defined relative to an external (from the respondent and the survey) truth, that being the views of experts who design courses to train students in the ways of thinking about and doing experimental physics. We highlight this here as the SEE score we are introducing is designed to identify a different type of truth, one that is internal to the respondent.

A similar approach to generating the YOU score from the YOU question can be used to score the Expert question. Both Zwickl~\cite{Zwickl2014} and Wilcox~\cite{Wilcox2017studentsviews} have compared the Expert and YOU scores to identify items where --- when comparing class/group average scores --- these differ significantly. They showed that respondents are generally aware of experts' views while not holding them themself in their lab courses. In our data (see Section~\ref{sec:data-source}), which includes some of the same data as from Ref.~\cite{Wilcox2017studentsviews}, we see a similar difference (Fig.~\ref{fig:eclass_violins}).

\subsubsection{Calculation of the SEE score}
The SEE score is calculated per item on the \eclass survey. The five-point Likert scale used for both the YOU and Expert questions is kept as numerical values from 1 to 5. Given that some items are phrased to elicit negative (Disagree) responses from experts, the scale of these items is reversed, such that for all items a higher value indicates agreement with the externally defined expert response, and is therefore similar to assigning $+1$ as used by the YOU score described in Section~\ref{sec:scoring-eclass} above.

The value for the Expert question response is then subtracted from the YOU response to give an integer value in the interval [-4,4], which is the SEE score for that response for that item. Therefore, the SEE score for an item has a 9-point scale. Similarly to the \eclass YOU score, we can sum over all items for a given respondent to find the total SEE score for an individual, which lies in the interval [-120,120]. 

\subsubsection{Interpretation and relation to student physics identity}\label{sec:methods:interpretation}

A SEE score of zero for an item would indicate that a student identifies themself as agreeing with a statement to the same extent that they believe an expert would agree with the statement. Therefore, we infer that they see themself as aligned with an expert view --- and thus identify in that item as an experimental physicist. 

SEE scores that are not zero indicate that the respondent positions themself as less-expert like, given they identify a difference between their own view and what they perceive an expert would respond. Negative SEE scores indicate the respondent's perception of how an expert would respond agrees with the consensus expert response (see Ref.~\cite{Zwickl2014}), while their own view is less in agreement. A positive score may occur when, for example, a respondent indicates they believe an expert would ``Agree'' with a statement (and that is consistent with the expert consensus), while their own view is that they ``Strongly Agree'' with the statement. This still indicates that the respondent does not share an expert-like view and is, therefore, important that the SEE score captures it. 

Another possible explanation for positive SEE scores, which we will find is more common, occurs when the perceived expert response to the statement is not aligned with the consensus expert response. For example, the respondent believes an expert would disagree with a statement, while the consensus response is that experts agree with the statement. This would then put the respondent's Expert response at a lower score (on the 1-5 scale) than their own response, resulting in a positive SEE score. This demonstrates one strength of the SEE score in not only identifying how far students see themselves from an expert, but also clearly indicating through a change in sign whether or not their perception of what an expert would say agrees with the consensus. In all of these cases, the respondent does not see themself as an expert, and hence there is a lack of self-recognition~\cite{hyater-adams_critical_2018}.

\subsubsection{Limitations of the \See score as a measure of self-recognition}\label{sec:limitations}

Given the wording of the item prompts (see Section~\ref{sec:methods:eclass-intro}), to interpret the results as a measure of identity, we have to assume that the goals of the lab course are aligned with those of expert experimental physicists. That is, what students are expected to do in their lab course is also what expert experimental physicists would expect to do when they are in a research lab. Indeed, in our analysis in Section~\ref{sec:results:pre-test-scores}, we will see that the \see score is able to identify when this assumption breaks down.

For the \see score to be valid, we also have to assume that respondents will score both YOU and Expert responses on the same scale for each item. Given the presentation of these questions (see Figure~\ref{fig:example-item}) immediately follow each other within the same item prompt and the response scales are aligned, we believe this is a reasonable assumption to make. Additionally, as the scales are numerically labelled rather than using text descriptors, this helps to justify interpreting the difference between the Expert and YOU responses as an immutable scale for each individual respondent. That is, for any given respondent, the difference between a score of 4 and 5 for YOU and Expert responses (SEE score of -1) is the same as the difference between a YOU and Expert responses of 1 and 2 (SEE score of -1 too). 

In generating the total SEE score, which sums over all E-CLASS items, there is the further assumption that the above scales remain the same between items. This, we again believe to be a reasonable assumption, based on the common presentation of the items and that the survey is completed in a single sitting of approximately 10--15 minutes. 

Comparing SEE scores pre-and post-test, we also assume that there is consistency in these scales for the respondents, despite there being a longer duration (typically months) between taking the surveys. We note that there is the added issue of response-shift bias, where student understanding of a concept at the post-test stage has changed such that the scale on which they grade, for example, their own self-efficacy, changes from pre-to-post test~\cite{miller_response-shift_2023}. This is, perhaps, one advantage of the SEE score as constructed, as each item effectively asks for the respondent to indicate their own comparator (the Expert response). Consequently, even if the respondent's own perception of the expert view changes during the course that is captured within the SEE score.   

As we will be considering aggregate data from multiple students on different courses at different institutions, when combining this data, we have to assume that there is an equivalence between the same SEE score values. That is a SEE score of -1 for one student means the same thing as it means for another student. This is a major assumption of this work. Normally, such fine differences are amalgamated using a reduced scale, such as described in Section~\ref{sec:scoring-eclass} for the YOU score. We argue that this assumption can be justified as each point on the scale from 0 to $\pm4$ indicates a distinct statement of positionality with respect to an expert view. Due to the symmetry of the situation, we only consider one side of the scale here. Clearly -4 and 0 are the extreme ends of this scale and indicate complete disagreement and agreement respectively. Then, -3 and -1, are both statements that are ``near'' either of those extremes. Finally, this leaves a score of -2, which shows hesitancy about one's own position, that of the expert view, or both. 

A final point to note is that the \eclass was designed to assess at the class-level rather than individual students~\cite{Zwickl2014}. Therefore, constructing a measure of identity using the data may seem counter-intuitive. However, we are still only proposing to use the measure on the class-level or demographic-group scale, much like the YOU score has been used previously, to look at trends in identities of students as expert physicists across time and within and between groups. 

\subsubsection{Practicalities of calculating the SEE score}
In calculating the SEE score, it is necessary to identify individuals who have not responded to the survey as intended by the survey designers. That is, respondents who select the same response for both the YOU and Expert questions for each and every item. We detail our approach to filtering out these responses in Appendix~\ref{app:data-cleaning}. We note that this does not affect historical analysis of \eclass data because that used only the responses to the YOU question, and it may be assumed that as this question is asked first, those responses may be interpreted as genuine.

\subsection{Demonstrating the distinction between YOU and SEE scores}

\subsubsection{Data sources}\label{sec:data-source}
We use a public data set consisting of more than 70,000 student responses to the E-CLASS survey~\cite{Aiken21} to analyse the constructed SEE score and how it differs from the YOU score. This data set consists of data from the E-CLASS survey collected between 2016 and 2019 and corresponds to 494 unique courses across 120 institutions, mainly located in the United States of America. For illustration purposes, in the first part of the results we present the distributions and basic statistics of the total SEE score.

\subsubsection{Correlation properties of total scores}\label{sec:methods:correlation-properties}
To establish that the SEE score provides different and hence new information compared to the YOU score, we first identify the correlation between each respondent's total scores on the two measures. The higher the correlation between the two measures the more shared information they have and the less extra insight that can be provided by calculating the SEE score.

We then consider the correlation between mean item scores of the SEE and YOU scores. In this case, by looking at individual items, we can identify clear outliers that indicate SEE and YOU scores measure different constructs.

\subsubsection{Temporal properties}\label{sec:methods:prepost}
Finally, we investigate how the SEE and YOU scores change over time by doing an item-wise pre-test/post-test comparison. The purpose here is to show that items can behave differently according to the two measures and, therefore, illustrate that the SEE and YOU scores are effectively measuring different quantities. To do this, we calculate Cohen's d for both the SEE and YOU scores for item $i$, as
\begin{equation}
    d_i = \frac{\mathrm{Post}_i - \mathrm{Pre}_i}{\sigma_{i,\mathrm{pooled}}},
\end{equation}
where $\mathrm{Post}_i$ is the mean over all students post-test score for item $i$, $\mathrm{Pre}_i$ is the mean over all students pre-test score for item $i$, and $\sigma_{i,\mathrm{pooled}}$ is the pooled standard deviation for item $i$, defined by Cohen~\cite[pg. 67, eq. 2.5.2]{cohen_statistical_1988}. We also estimate the standard error on the calculated Cohen's d values using Borenstein and Hedge's method~\cite{Borenstein2019}. We use Cohen's d for this comparison because the SEE and YOU scores have different scales, and Cohen's d, as an effect size, normalises these differences and therefore facilitates comparisons between the two scores.

\section{Quantitative Results and Discussion}\label{sec:results}

\subsection{Basic properties of \see scores}
\begin{figure}
    \centering
    \includegraphics[width=\linewidth]{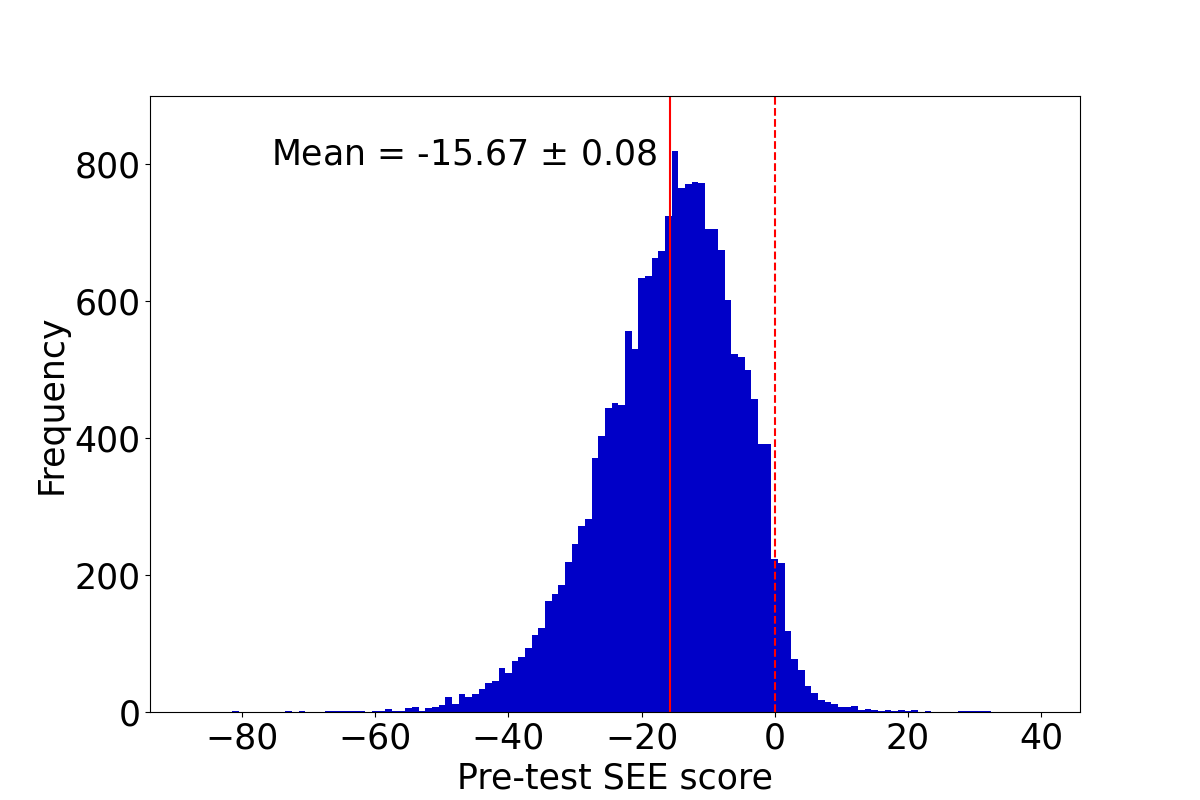}
    \caption{Distribution of the pre-test \see score. Vertical solid red line indicates the mean value of $-15.67\pm0.08$.}
    \label{fig:pre-dist}
\end{figure}

\begin{figure}
    \centering
    \includegraphics[width=\linewidth]{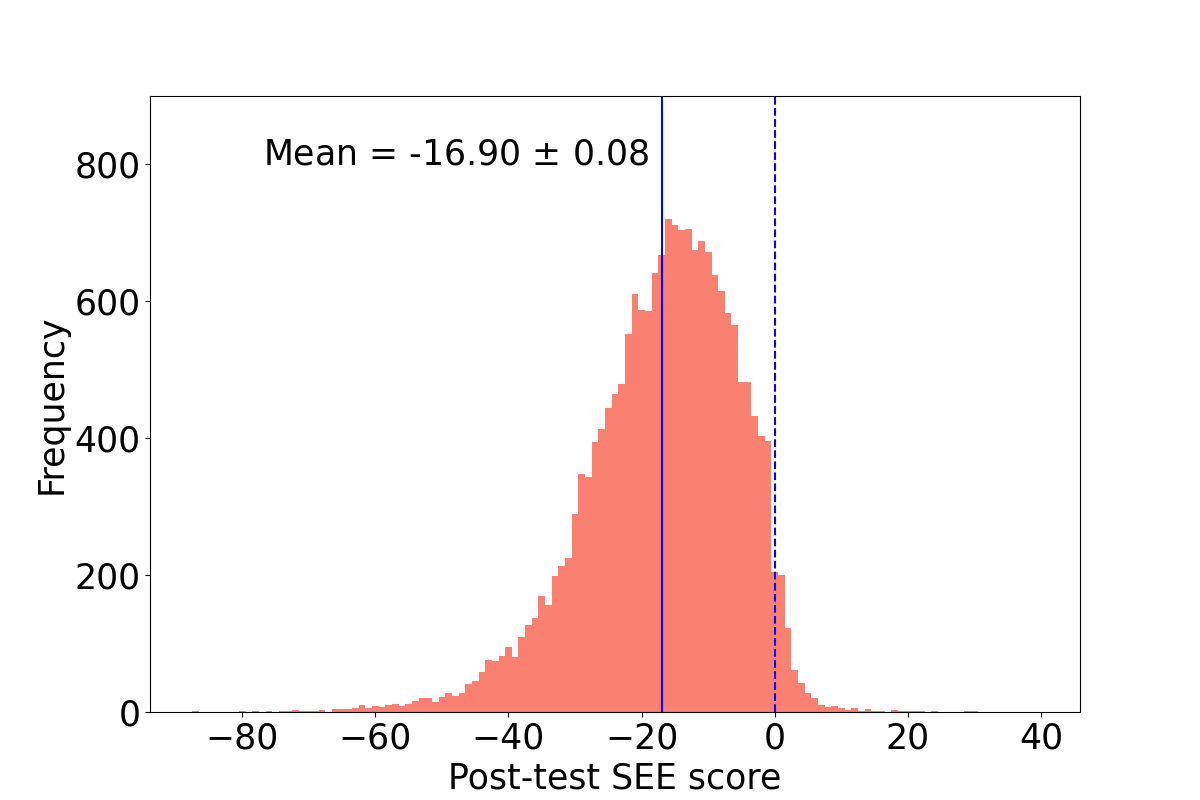}
    \caption{Distribution of the post-test \see score. Vertical solid red line indicates the mean value of $-16.90\pm0.08$.}
    \label{fig:post-dist}
\end{figure}

The mean of the total SEE score across all students for the pre-test is $-15.67\pm0.08$, therefore indicating that on average students position themselves approximately 0.5 points on the 5-point scale below what they view an expert experimental physicist would respond. This suggests that most students hold views close to, but not exactly matching, expert-like views (c.f., Fig.~\ref{fig:eclass_violins}). This contributes to our argument for maintaining the fine response scale in the SEE score.

The distribution for the post-test scores is similar to that for the pre-test score (compare Fig.~\ref{fig:pre-dist} and Fig.~\ref{fig:post-dist}), but with a lower mean total SEE score of $-16.90\pm0.08$. The negative shift from pre-test to post-test is consistent with the negative shift seen in YOU scores~\cite{wilcox2018summary}. Here, we interpret the negative shift through the identity lens rather than simply whether or not students hold expert-like views about their teaching labs. That is, we interpret this negative shift as students, on average, recognizing themselves less as expert experimental physicists and hence having less identity as a ``physics person''. 

The distribution of total SEE scores around the mean vary from $-81$ to $+32$ for the pre-test and $-87$ to $+30$ for the post-test. These high magnitude negative scores are extreme tails of the distribution, with 90\% of total scores between $-34$ and $-1$ for the pre-test and $-37$ and $-1$ for the post-test (5th to 95th percentile). This supports the idea that the most common size of difference between YOU and Expert responses is at most 1 point on the survey item scale. There are comparatively very few students with positive total scores (641 students in the pre-test and 538 students in the post-test; less than 3.4\% of all students).

In addition to an overall shift to more negative values of the SEE score in the post test (median decreases from $-15$ to $-16$), there is a higher negative skew in the post-test distribution (skew $= -0.75$) than in the pre-test distribution (skew $= -0.52$), suggesting more extreme values for the total SEE score occur in the post-test.

\subsection{Correlation between \See and YOU scores}
The Spearman rank-order correlation coefficient between YOU and SEE pre-test total scores is $0.51\pm0.01$. The value is not near unity, suggesting that the YOU and SEE scores are not identical. 

We expect that the YOU pre-test and post-test scores to have some correlation. The Spearman rank correlation coefficient between these two is $0.62\pm0.01$. We note that the YOU and SEE pre-test correlation value is lower than the YOU pre-test and post-test correlation, despite the YOU and SEE pre-test data sharing half of the data points. This provides evidence that the SEE score is measuring a different quantity than the YOU score.

\subsection{Comparing item-wise E-CLASS YOU and SEE pre-test scores}\label{sec:results:pre-test-scores}
\begin{figure*}[th]
    \centering
    \includegraphics[width=0.67\linewidth]{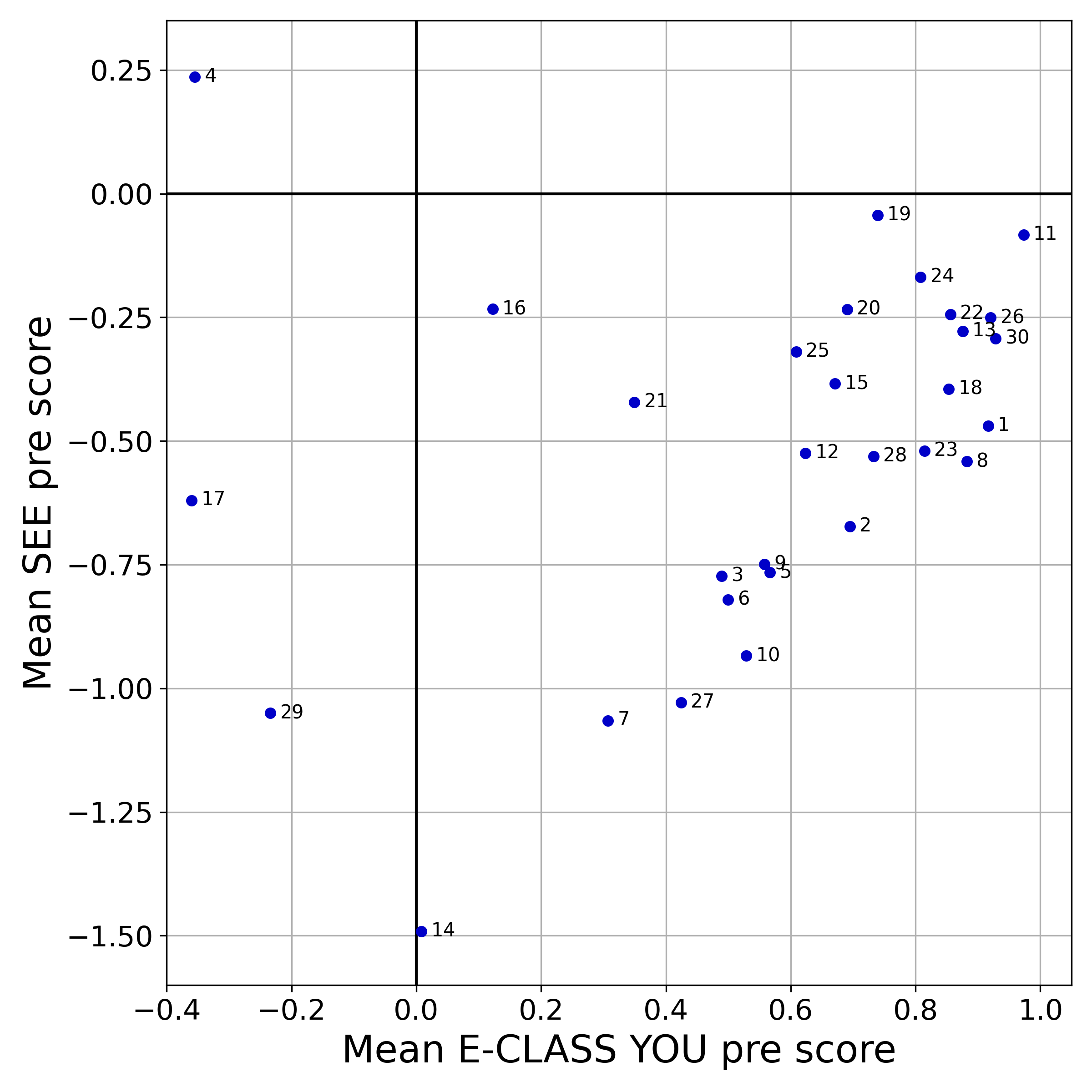}
    \caption{Plot of item mean SEE scores against mean YOU scores in the pre-test. Numbers indicate the question number of the data point. Error bars are not shown for clarity and also because they are smaller than the marker size; the uncertainty on each point is reported in Table~\ref{tab:B1_table}.}
    \label{fig:you-see-pre}
\end{figure*}

In this section, we first consider the overall relation between YOU and SEE pre-test scores, then consider specific items to illustrate how YOU and SEE scores differ. 

There is a positive Spearman rank-order correlation between the item-averaged \eclass YOU pre-test scores and SEE pre-test scores of $0.47\pm0.30$. If these two were highly correlated then we could conclude that the SEE score does not provide additional information. Considering Figure~\ref{fig:you-see-pre}, there appears to be a general clustering with a positive relationship from item 14 through to item 11. An explanation for this trend is that the SEE score is mainly reflecting the YOU score, that is the Expert score for these items does not tend to vary. The handful of items that do not belong to this cluster contribute to the large uncertainty on the calculated correlation value.

To orient ourselves on Figure~\ref{fig:you-see-pre}, let us first consider negative SEE scores close to zero. These are scores that correspond to situations when the average respondent identifies their own views as close to expert views. For these SEE scores, a YOU score close to unity indicates the respondent's YOU and Expert views and the consensus view are in alignment. In contrast, item 16, ``The primary purpose of doing a physics experiment is to confirm previously known results'', which has YOU score close to zero ($0.122\pm0.006$) and a small SEE score of $-0.233\pm0.006$, shows that while respondents only just disagree with the statement (the consensus expert view), they also tend to believe experts hold a similar view. Thus we may deduce that these respondents mistakenly identify themselves as holding expert-like views. This illustrates that the SEE score is telling us something different from the YOU score, in that these respondents see themselves as experts by holding such a view. This also then has implications for instructional strategies to develop those views, given the inertia associated with them being part of the respondent's identity.

In Figure~\ref{fig:you-see-pre}, item 4 is the most stark, as it has a positive SEE score ($0.236\pm0.007$). Our interpretation of this is as follows: the negative YOU score ($-0.355\pm0.006$) indicates that respondents' view on this item disagrees with the consensus expert view. The item states: ``If I am communicating results from an experiment, my main goal is to have the correct sections and formatting'' and the consensus expert view was to disagree with that statement~\cite{Zwickl2014}. Therefore, we know that students tend to agree with that statement. If respondents believe experts think it is important, then they may position themselves further away from the perceived expert view by, for example, selecting `Agree' for themself in the YOU response and `Strongly Agree' for the expert response. As a result of this mis-alignment of the perceived expert view with the consensus expert view, then the students have actually positioned themselves closer to the consensus expert view, which will result in a positive SEE score. Here, we see that while the YOU score would identify that students disagree with experts, the SEE score adds extra information that this is actually what they think experts believe. We will return to this item in Section~\ref{sec:results:pre-post-shifts} to see what the result of instruction is on this view point.

Item 17 also has a negative YOU score ($-0.360\pm0.006$), indicating respondents agree with the statement: ``When I encounter difficulties in the lab, my first step is to ask an expert, like the instructor'' while the consensus expert response is to disagree with it. From the SEE score ($-0.620\pm0.008$), we find that respondents regularly position their view just over half a Likert scale within their expectation of how an expert would answer. Comparing with item 4, which has a similar YOU score, shows how the SEE score provides clear differentiation between these items in that for item 17 respondents (correctly) position themself further from the expert-like view than item 4. 

From the above examples, we see that the SEE score clearly shows different properties to the YOU score. We have not considered the post-test scores, where it would be possible to perform a similar comparison, because in the next section we focus on the pre-to-post shifts, which utilises the post-test information and therefore would make such an analysis redundant.

\subsection{Comparing item-wise E-CLASS YOU and SEE pre-post shifts}\label{sec:results:pre-post-shifts}
\begin{figure*}[th]
    \centering
    \includegraphics[width=0.67\linewidth]{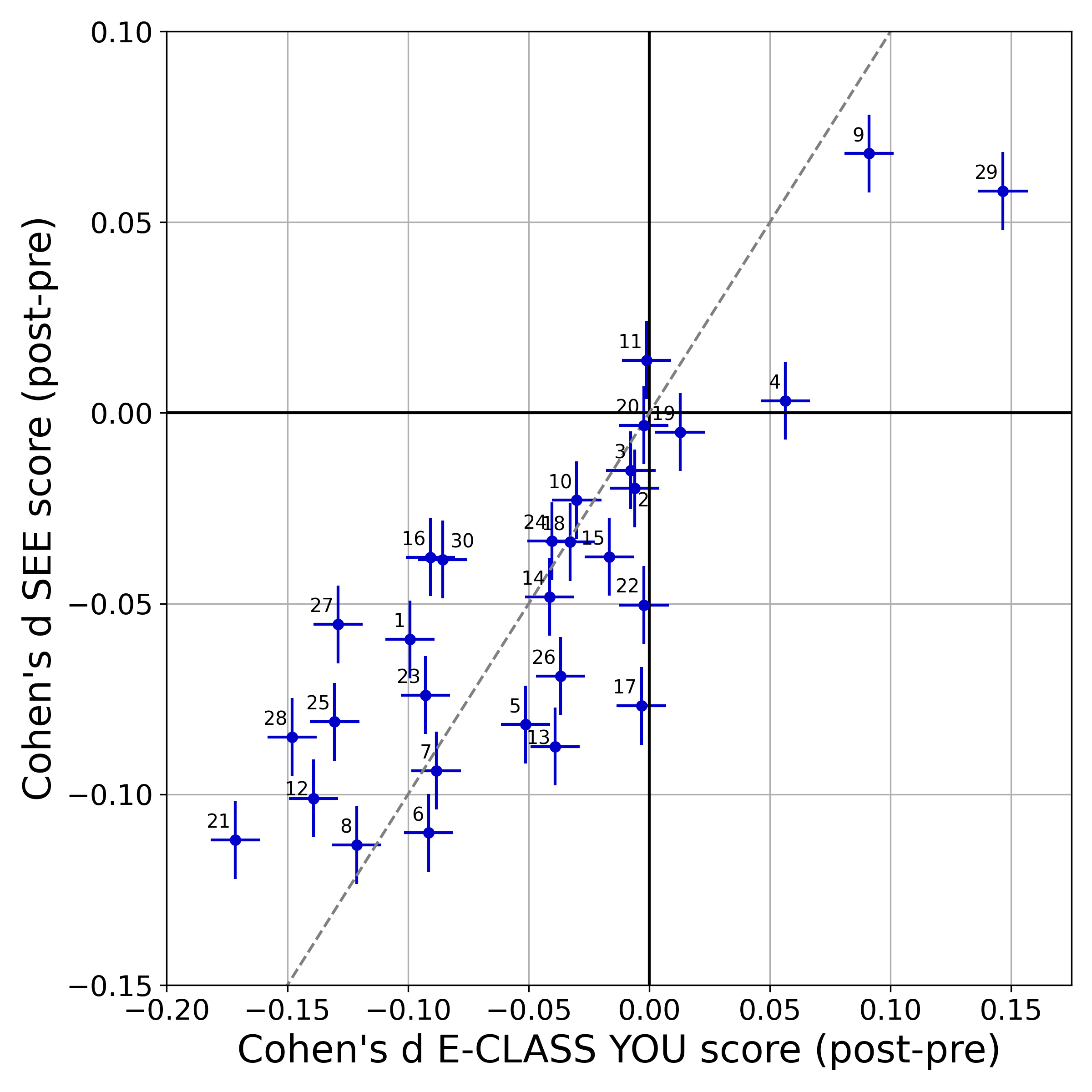}
    \caption{Plot of the change from pre-test to post-test as measured by Cohen's d in the item mean SEE scores against the change in mean YOU scores. Numbers indicate the question number of the data point. Error bars indicate an estimate of the sample variance on Cohen's d~\cite{Borenstein2019}. These data are derived from the values given in Table~\ref{tab:B1_table}.}
    \label{fig:you-see-delta}
\end{figure*}

We now look at the effect of instruction on the SEE score in comparison with the effect on the YOU score. \eclass YOU scores on average drop from pre-test to post-test for the majority of courses~\cite{Wilcox18,Aiken21}. The SEE score also shows a negative shift from pre-test to post-test, and has a strong correlation with the YOU score, see Figure~\ref{fig:you-see-delta} (Spearman rank-order correlation $0.808\pm0.006$). Generally, we can say that for most items students position themselves further away from an expert at the end of the course than at the beginning. This could be due three possible reasons: students' own views become less expert-like, with their idea of what experts think remaining constant; the opposite, where students' own views stay the same, while they have changed their views of what experts think; or both their own views and their views of experts have changed. In this section, we do not elaborate in detail which of these three possibilities occurs for the items we discuss, because by construction, the SEE score does not distinguish between them. The SEE score has been designed to be a measure of self-recognition and, therefore, to help identify to what extent students associate their own practices and beliefs with their perception of experts' practices and beliefs. Nevertheless, there is also value for instructors in knowing which of the three explanations occurs for each item, as, for example, if it is students' views of experts' views that changes, then instructors may be better informed when making changes to the course. Therefore, for completeness, we include a brief analysis of this in Appendix~\ref{app:expert-shift}. In this Appendix, we see that for 8 items the change in Expert score is greater than the change in YOU score, showing that while the SEE and YOU pre-to-post changes are highly correlated, this correlation does not imply that the change is only coming from the YOU component of the SEE score.

For the purposes of this current work in demonstrating that SEE and YOU provide different information, we focus now in our discussion of these results on the items that show clear changes in one measure and not the other: 
items 4 and 17, both discussed in Section~\ref{sec:results:pre-test-scores}.

As a reminder, item 4, states: ``If I am communicating results from an experiment, my main goal is to have the correct sections and formatting.'' The positive YOU score shift indicates that respondents hold more expert-like views after instruction than before. However, the mean SEE score is not different from zero, therefore, suggesting that the Expert view has also changed (see Fig.~\ref{fig:post-pre-expert-you} in Appendix~\ref{app:expert-shift}). In interpreting the SEE score as a measure of self-recognition, we see there has not been a quantitative change, despite there being a qualitative change in the how respondents understand what it means to be an expert. As we are claiming that the SEE score measures self-recognition, it makes sense that if both yardsticks move then no change should be seen in this measure (i.e., we are seeing response-shift bias by design).

In contrast, item 17 (``When I encounter difficulties in the lab, my first step is to ask an expert, like the instructor.'') shows no change in YOU score from pre-test to post-test, while having a decrease in SEE score. Therefore, respondents are positioning themself further away from an expert view by re-evaluating their perception of how experts would respond to this item statement, while not changing their own view. This is perhaps the strongest evidence that the SEE score practically measures a different quantity than YOU scores. 

We note that, with the data presented, all of the pre-to-post shifts for each item (Fig.~\ref{fig:you-see-delta}) showed negligible effect sizes (Cohen's $d < 0.2$~\cite{fritz2012effect}). Indeed, the changes in YOU score also showed similar effect sizes. Nevertheless, we may still draw conclusions based on the observed changes due to the large number of responses being analysed and, hence, the low uncertainty in each data point. 

One possible reason that we do not see larger effect sizes is that we are performing this analysis by averaging over a large number of students completing experimental physics courses at multiple levels (introductory and beyond-first year) and at 120 different institutions. This averaging deliberately ignores the variation between the contexts and course learning goals and acts to wash out larger changes in specific contexts or for specific demographic groups. This is appropriate for the current work because we are interested only in demonstrating the existence of a difference between the newly constructed SEE score and the YOU score. The differences seen in the scores for items that we have chosen are potentially common across many courses, giving a view into the landscape of undergraduate lab courses in the USA (see Ref.~\cite{Holmes2020landscape,Geschwind2024}), but further studies would be required to identify how the SEE score, and this one component of student physics identity, varies for individual courses.

\section{Conclusion}
We have presented a new measure, the Self-Evaluated Expertise (SEE) score, that fully utilises the data collected during the normal implementation of the E-CLASS by combining both the YOU and Expert responses to the survey items. In this work, we argue that this construction provides a measure of self-recognition within the physics identity framework of Hyater-Adams et al.~\cite{hyater-adams_critical_2018} and, therefore, can provide insights into the extent to which students view themselves as experts. Naturally, this measure has numerous limitations in how it should be interpreted, which we have highlighted in Section~\ref{sec:limitations}.

We have illustrated using open-source data~\cite{Aiken21} that the SEE score provides different information to that of the traditional E-CLASS scoring system (YOU score), in both a pre-test comparison and in pre-test to post-test analysis. This demonstrated that while there is a medium correlation between YOU and SEE scores, there are distinct cases where the SEE score changes and the YOU score does not and vice versa. 

The theoretical construction of Self-Evaluate Expertise as a measure of student physics identity allows us to conclude that  the decrease in SEE score from pre-test to post-test indicates that on average (across courses and institutions) students' self-recognition decreases and in ceteris paribus so too does their physics identity. This has clear implications for retention of students within the physics discipline~\cite{Cwik2022recognized,Rosenberg2024mentorship,Brew2025persistence}.

The SEE score we hope will be of interest to instructors, who might have aims for their course to develop student identity as an experimental physicist, and also to researchers looking at how different course contexts may help or hinder the development of physics identities. 

\section{Acknowledgements}
We would like to thank the Imperial College London Department of Physics Undergraduate Research Opportunities Programme for funding TP's work on this project. 

\bibliographystyle{unsrtnat}
\bibliography{annot}

\clearpage
\newpage
\onecolumngrid
\appendix

\section{Respondent demographic and course details} \label{sec:demographics}

\begin{table*}[th]
    \centering
    \renewcommand{\arraystretch}{1.2}
    \begin{tabular}{llllll} \toprule
    \multicolumn{3}{r}{\emph{Gender}}&\\\cmidrule{2-5}
    {\emph{Major}} & {Undeclared} & {Women} & {Men} & {GNC} & {\emph{Total}} \\\cmidrule{1-6}
    Non-Science & 63 & 692 & 1839 & 45 & 2639 \\
    Other Science & 162 & 4386 & 3835 & 61 & 8444 \\
    Engineering & 83 & 1897 & 3581 & 64 & 5625 \\
    Physics & 102 & 1011 & 1361 & 27 & 2501 \\
    \\
    {\emph{Course Type}}&{}&{}&{}&{}&{\emph{Total}}\\\midrule
    First Year/Introductory & 342 & 7106 & 8907 & 153 & 16508 \\
    Beyond First Year & 68 & 880 & 1709 & 44 & 2701 \\
    \\
    {\emph{Ethnicicty}}&{}&{}&{}&{}&{\emph{Total}}\\\midrule
    Undeclared & 295 & 374 & 450 & 15 & 1134 \\
    American Indian or Alaska Native & 0 & 20 & 25 & 0 & 45 \\
    Asian &  42 & 1570 & 2044 & 21 & 3677 \\
    Black or African American & 5 & 539 & 396 & 8 & 948 \\
    Hispanic/Latino & 9 & 479 & 793 & 10 & 1291 \\
    Native Hawaiian or other Pacific Islander & 1 & 16 & 40 & 0 & 57 \\
    White  & 44 & 4240 & 5885 & 86 & 10255 \\
    Other & 2 & 142 & 204 & 34 & 382 \\
    More than one ethnicity & 12 & 606 & 779 & 23 & 1420 \\\midrule
    \emph{Total} & 410 & 7986 & 10616 & 197 & \boldsymbol{\mathbf{19209}}\ \\\bottomrule
    \end{tabular}
    \caption{Demographic and course detail breakdown collected from the post-test responses and instructor course information survey respectively. (GNC = gender non-conforming).}
    \label{tab:demographic_table}
\end{table*}
\noindent
The instructor provided details of the course and in the post-test respondents were asked to provide demographic details (Table \ref{tab:demographic_table}). Of all the respondents: 42\% were women, 55\% were men, 2\% were gender non-conforming, and 1\% did not declare their gender. 

Of the students who completed the E-CLASS in our sample: 13\% were physics majors (which includes students who major in physics, engineering physics, or astrophysics); 29\% were engineering majors; 44\% were majoring in ``other" sciences (chemistry, biochemistry, biology, astronomy, geology, mathematics, computer science, physiology); and lastly 14\% of students either did not declare their major or were majoring in a ``non-science" subject. 86\% of the matched E-CLASS responses were from students taking introductory courses, while the remaining 14\% of the respondents were completing a beyond-first-year course. 

The results of the post-instruction E-CLASS survey also reveal insights into the ethnic breakdown of the respondents. 87\% declared themselves as belonging to exactly one ethnic group, 6\% of respondents chose not to declare their ethnic group, and 7\% of respondents declared themselves as belonging to more than one ethnic group. Out of all respondents, 0.2\% declared themselves as ``American Indian or Alaska Native", 19.1\% declared themselves as ``Asian", 4.9\% declared themselves as ``Black or African American", 6.7\% declared themselves as ``Hispanic/Latino", 0.3\% declared themselves as ``Native Hawaiian or other Pacific Islander", 53.4\% declared themselves as ``White", and 2\% declared themselves as having another ethnicity that was not listed as an option under provided groups.

\clearpage

\section{Data cleaning}\label{app:data-cleaning}
\begin{figure*}[ht]
    \centering
    \includegraphics[width=0.8\linewidth]{"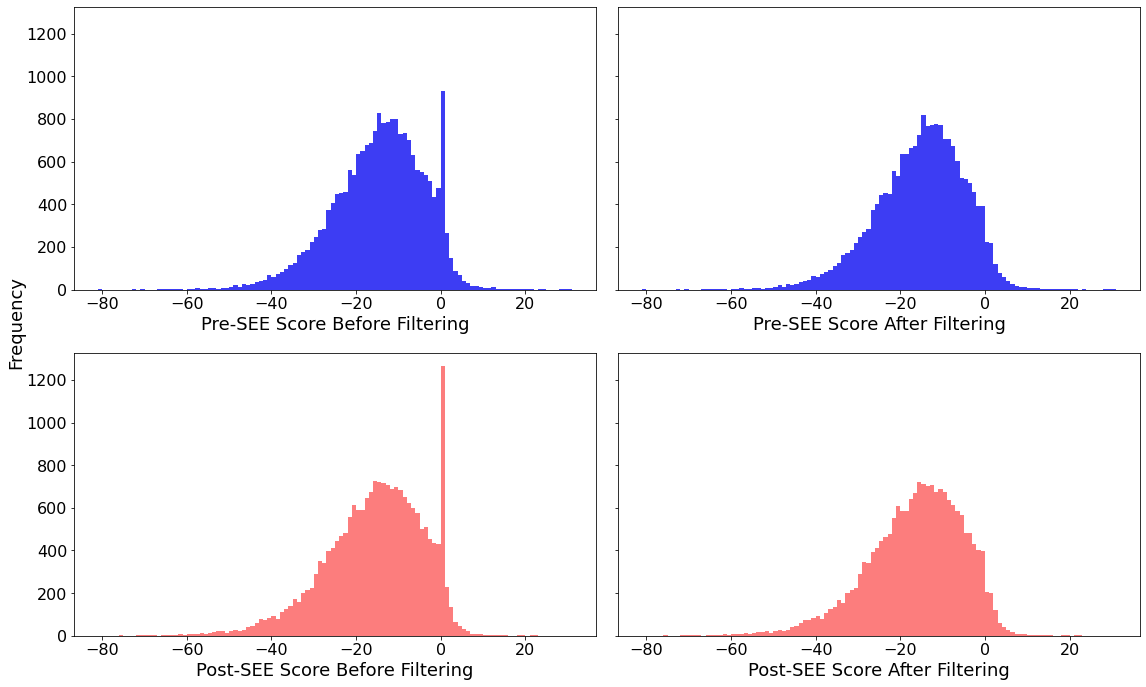"}
    \caption{Histogram plots depicting distribution of total SEE scores, for the pre-test and post-test, before and after filtering out invalid responses at the origin, where students are likely to have misunderstood the survey.}
    \label{fig:SEE_dist}
\end{figure*}
The distribution of SEE scores also revealed an insight into the validity of responses to the E-CLASS survey. After plotting the distribution of the SEE scores pre-test and post-test: the distributions returned an observable peak of SEE values at the origin. This observation can be seen in figure \ref{fig:SEE_dist}. An overall SEE score of 0 implies that a student's view of experimental physics does not differ to that of the student's own view. Obtaining a SEE score of 0 can be done in one of two ways. Either: for each item, the student had non-zero SEE scores, yet, the student cumulatively obtained 0 as an overall SEE score; or for each item, the student obtained a SEE score of 0 by selecting the same response for the YOU item and Expert item, for all 30 questions to the survey. In the case of the latter occurrence, it is likely that the student misunderstood the survey: and so the responses whose SEE score for each item was equal to zero, either pre-test or post-test, were subsequently filtered out. Before filtering, there were 20,617 matched responses in the E-CLASS data-set. It was found that 1,408 responses, pre-test or post-test, had a SEE score of 0 for each item on the survey. Filtering these invalid responses out of the survey reduced the overall sample size of the data set to 19,209 responses. The histogram plots demonstrate the effect of removing these invalid responses from the survey. It creates a much more even distribution of SEE scores whereby the distribution of SEE scores is no longer discontinuous at 0: thus validating their removal. 

Understandably the mean of the SEE scores in the pre-test and post-test became more negative following filtering. The mean of the SEE scores before filtering was -14.85 $\pm$ 0.07 in the pre-test and -15.87 $\pm$ 0.08 in the post-test. After filtering, the mean of the SEE scores was -15.67 $\pm$ 0.08 in the pre-test and -16.90 $\pm$ 0.08 in the post-test. 

\clearpage

\section{Mean scores for survey items}\label{app:data}
\begin{table}[ht]
    \centering
        \caption{Mean and standard errors for E-CLASS and \See pre-test and post-test scores ordered by increasing E-CLASS YOU pre-test score.}\label{tab:B1_table} 
    \renewcommand{\arraystretch}{1}
    \begin{tabular}{rcccc} \toprule
    {} &
   \multicolumn{2}{c}{E-CLASS YOU}&\multicolumn{2}{c}{\See}\\ 
   \cmidrule{2-5}
   {Question Number} & {Pre} & {Post} & {Pre} & {Post}\\ 
   \midrule
17              & $-0.360 \pm 0.006$      & $-0.363 \pm 0.006$       & $-0.620 \pm 0.008$      & $-0.704 \pm 0.008$       \\
4               & $-0.355 \pm 0.006$      & $-0.309 \pm 0.006$       & $0.236 \pm 0.007$       & $0.238 \pm 0.007$        \\
29              & $-0.234 \pm 0.006$      & $-0.110 \pm 0.006$       & $-1.050 \pm 0.008$      & $-0.986 \pm 0.008$       \\
14              & $0.008 \pm 0.006$       & $-0.026 \pm 0.006$       & $-1.492 \pm 0.008$      & $-1.546 \pm 0.008$       \\
16              & $0.122 \pm 0.006$       & $0.048 \pm 0.006$        & $-0.233 \pm 0.006$      & $-0.268 \pm 0.006$       \\
7               & $0.307 \pm 0.006$       & $0.237 \pm 0.006$        & $-1.065 \pm 0.009$      & $-1.187 \pm 0.010$       \\
21              & $0.349 \pm 0.006$       & $0.203 \pm 0.006$        & $-0.421 \pm 0.008$      & $-0.542 \pm 0.008$       \\
27              & $0.424 \pm 0.006$       & $0.321 \pm 0.006$        & $-1.029 \pm 0.007$      & $-1.085 \pm 0.007$       \\
3               & $0.489 \pm 0.005$       & $0.483 \pm 0.006$        & $-0.773 \pm 0.008$      & $-0.789 \pm 0.008$       \\
6               & $0.500 \pm 0.005$       & $0.434 \pm 0.005$        & $-0.821 \pm 0.007$      & $-0.935 \pm 0.008$       \\
10              & $0.529 \pm 0.005$       & $0.507 \pm 0.005$        & $-0.934 \pm 0.007$      & $-0.956 \pm 0.007$       \\
9               & $0.558 \pm 0.005$       & $0.621 \pm 0.005$        & $-0.749 \pm 0.007$      & $-0.688 \pm 0.006$       \\
5               & $0.567 \pm 0.005$       & $0.531 \pm 0.005$        & $-0.766 \pm 0.006$      & $-0.840 \pm 0.007$       \\
25              & $0.609 \pm 0.005$       & $0.513 \pm 0.006$        & $-0.319 \pm 0.006$      & $-0.383 \pm 0.006$       \\
12              & $0.624 \pm 0.005$       & $0.520 \pm 0.006$        & $-0.525 \pm 0.006$      & $-0.612 \pm 0.006$       \\
15              & $0.671 \pm 0.004$       & $0.661 \pm 0.004$        & $-0.384 \pm 0.005$      & $-0.412 \pm 0.005$       \\
20              & $0.690 \pm 0.004$       & $0.689 \pm 0.005$        & $-0.234 \pm 0.008$      & $-0.237 \pm 0.008$       \\
2               & $0.695 \pm 0.004$       & $0.691 \pm 0.004$        & $-0.673 \pm 0.006$      & $-0.691 \pm 0.007$       \\
28              & $0.732 \pm 0.004$       & $0.639 \pm 0.005$        & $-0.531 \pm 0.006$      & $-0.605 \pm 0.007$       \\
19              & $0.739 \pm 0.004$       & $0.746 \pm 0.004$        & $-0.044 \pm 0.006$      & $-0.048 \pm 0.006$       \\
24              & $0.808 \pm 0.004$       & $0.787 \pm 0.004$        & $-0.169 \pm 0.005$      & $-0.192 \pm 0.005$       \\
23              & $0.814 \pm 0.004$       & $0.766 \pm 0.004$        & $-0.520 \pm 0.006$      & $-0.579 \pm 0.006$       \\
18              & $0.853 \pm 0.003$       & $0.839 \pm 0.003$        & $-0.395 \pm 0.005$      & $-0.420 \pm 0.005$       \\
22              & $0.856 \pm 0.003$       & $0.855 \pm 0.003$        & $-0.244 \pm 0.004$      & $-0.275 \pm 0.005$       \\
13              & $0.876 \pm 0.003$       & $0.859 \pm 0.003$        & $-0.278 \pm 0.005$      & $-0.339 \pm 0.005$       \\
8               & $0.882 \pm 0.003$       & $0.829 \pm 0.003$        & $-0.541 \pm 0.005$      & $-0.627 \pm 0.006$       \\
1               & $0.916 \pm 0.002$       & $0.879 \pm 0.003$        & $-0.470 \pm 0.005$      & $-0.515 \pm 0.006$       \\
26              & $0.920 \pm 0.002$       & $0.909 \pm 0.002$        & $-0.251 \pm 0.004$      & $-0.288 \pm 0.004$       \\
30              & $0.928 \pm 0.002$       & $0.899 \pm 0.003$        & $-0.293 \pm 0.004$      & $-0.316 \pm 0.005$       \\
11              & $0.973 \pm 0.001$       & $0.973 \pm 0.001$        & $-0.083 \pm 0.003$      & $-0.077 \pm 0.003$      \\
\bottomrule
\end{tabular}
\end{table}

Table~\ref{tab:B1_table} shows the mean pre-test and post-test YOU and SEE scores for all items. These data are plotted in Figure~\ref{fig:you-see-pre} for the pre-test and the change (post minus pre) are plotted in Figure~\ref{fig:you-see-delta}. 

\clearpage

\section{YOU and Expert shifts from pre-test to post-test}\label{app:expert-shift}

\begin{figure}[h]
    \centering
    \includegraphics[width=\linewidth]{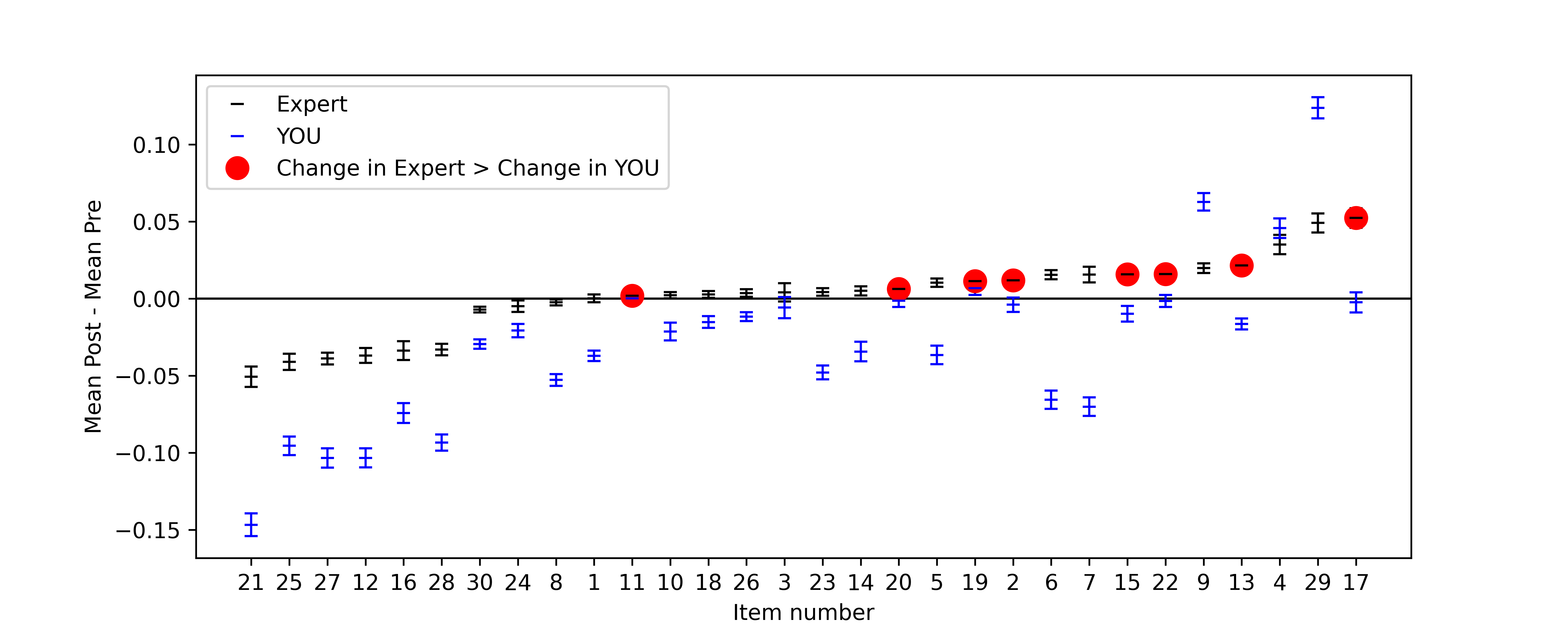}
    \caption{Plot of the difference between mean scores on the post-test minus the pre-test for Expert and YOU scores for each item. Items are ordered from most negative change in Expert score from pre-test to post-test to most positive change. Red circles identify the items where the Expert score change is greater than the YOU score change. Error bars show the standard error on the mean.}
    \label{fig:post-pre-expert-you}
\end{figure}

The SEE score is constructed from the difference between the YOU and the Expert response for each item for each student completing the E-CLASS survey. Therefore, the SEE score can change, as described in Section~\ref{sec:results:pre-post-shifts}, due to changes in either of these two quantities. In this Appendix, we provide the data on the changes for each item in Figure~\ref{fig:post-pre-expert-you} such that an interested reader may determine which component of the SEE score contributed to the change. 

\end{document}